\documentclass[12pt,osajnl,copyedit]{revtex4}  

\usepackage[dvips,colorlinks=true,bookmarks=false,citecolor=blue,
urlcolor=blue,pdfstartview=FitH]{hyperref}

\input{Qcircuit} 
\usepackage{graphicx}
\usepackage{epsfig}
\usepackage{dcolumn}
\usepackage{bm}
\usepackage[latin1]{inputenc}

\begin{document}

\title{Photonic entanglement as a resource in quantum computation and quantum communication}

\author{Robert Prevedel}
\affiliation{Institut für Experimentalphysik, Universit\"at Wien,
Boltzmanngasse 5, 1090 Wien, Austria}

\author{Markus Aspelmeyer}
\affiliation{Institut für Experimentalphysik, Universit\"at
Wien, Boltzmanngasse 5, 1090 Wien, Austria\\and\\Institut für Quantenoptik und Quanteninformation,\\
 \"Osterreichische Akademie der Wissenschaften, Boltzmanngasse 3, 1090 Wien, Austria}

\author{$\check{\emph{C}}$aslav Brukner}
\affiliation{Institut für Experimentalphysik, Universit\"at
Wien, Boltzmanngasse 5, 1090 Wien, Austria\\and\\Institut für Quantenoptik und Quanteninformation,\\
 \"Osterreichische Akademie der Wissenschaften, Boltzmanngasse 3, 1090 Wien, Austria}

\author{Thomas D. Jennewein}
\affiliation{Institut für Quantenoptik und Quanteninformation,\\
 \"Osterreichische Akademie der Wissenschaften, Boltzmanngasse 3, 1090 Wien, Austria}

\author{Anton Zeilinger}
\affiliation{Institut für Experimentalphysik, Universit\"at
Wien, Boltzmanngasse 5, 1090 Wien, Austria\\and\\Institut für Quantenoptik und Quanteninformation,\\
 \"Osterreichische Akademie der Wissenschaften, Boltzmanngasse 3, 1090 Wien, Austria}

\begin{abstract}Entanglement is an essential resource in current experimental
implementations for quantum information processing. We review a
class of experiments exploiting \textit{photonic }entanglement,
ranging from one-way quantum computing over quantum communication
complexity to long-distance quantum communication. We then propose
a set of feasible experiments that will underline the advantages
of photonic entanglement for quantum information processing.\\
\end{abstract}


\maketitle 

\section{Introduction}

\renewcommand{\baselinestretch}{2.0}\normalsize

Quantum entanglement~\cite{Schrodinger35a} has become an important
resource for many practical tasks in quantum information
processing such as quantum computing, quantum communication or
quantum metrology. From an early stage on, entanglement proved to
be an essential tool for quantum physics, both in theory and
experiment: early experimental realizations of entangled photon
pairs were used to demonstrate the quantum nature of polarization
correlations that can occur in decay
processes~\cite{Wu_1950,Kocher_1967}, to confirm quantum
predictions of radiation theory and falsify semi-classical
models~\cite{Clauser72,Clauser74}, or to test Bell' s theorem and
exclude local realistic descriptions of the observed quantum
phenomena~\cite{Bell64,Freedman72,Aspect82b,Weihs98}. It followed
the discovery of information processing for quantum physics (and
vice versa), partly triggered by the introduction of quantum
cryptography~\cite{Wiesner83,Bennett84,Ekert91}, and hence the
beginning of quantum information science, which has evolved to a
strongly expanding branch of science. Entanglement is a
fundamental resource for it, as a quantum channel in quantum
communication (e.g. for quantum state
teleportation~\cite{Bennett93,Bouwmeester97} or quantum dense
coding~\cite{Bennett92b,Mattle96}) or as computational resource.
Quantum computing with photons has recently experienced a new
boom by discovering the possibility of universal computing with
linear optics and measurements alone~\cite{Knill00}. Although it
is still unclear what the minimal resource requirements for
optical quantum computing are, the number of required optical
elements per universal gate is constantly decreasing. Another
appealing feature of photonic quantum computing is the possibility
of gate times much faster
than in any other physical implementation to date. \\
In the following we will discuss new examples involving
experiments on entangled photons that underline the importance of
entanglement for quantum information processing. Section 2 starts
with an introduction to photonic one-way quantum computing, a new
approach that makes optimal use of entanglement as a resource. We
propose an experiment to achieve deterministic quantum computing,
a unique feature of the one-way quantum computer, by introducing
active corrections during the computation. Section 3 describes
experimental challenges and perspectives when exploiting
distributed entanglement for quantum networking tasks, in
particular long-distance quantum communication,
higher-dimensional quantum cryptography and quantum communication
complexity.

\section{Towards deterministic One-Way Quantum Computing with active Feed-Forward}
Linear optical quantum computing (LOQC) is one of the promising
candidates for the physical realization of quantum computers. LOQC
employs photonic qubits as information carriers, which have the
immense advantage of suffering negligible decoherence and
providing high-speed gate operations. It was shown that linear
optics and projective measurements allow for essential nonlinear
interactions and eventually for scalable quantum
computing~\cite{Knill00}. This has led to a flurry of research in
both theory and experiments. A recent and comprehensive overview
can be found in~\cite{Kok_2005}. The intrinsic randomness of the
projective measurements in linear optics, however, only allows for
probabilistic gate operations, i.e., the gate operations are
successful only in a small fraction of the time. The other times
the outcomes need to be discarded. Although the gate success
probability increases with additional resources (optical elements
and/or ancilla photons), such schemes achieve nearly
deterministic gate operations only in the asymptotic regime of
infinite resources, which is experimentally infeasible. In
contrast, the one-way quantum computer model
\cite{Raussendorf01,Nielsen04b}, an exciting alternative approach
to LOQC, allows the resource for the quantum computation to be
prepared \emph{offline} prior to any logical operations. The
computational resource is a highly entangled state (the so-called
cluster state). Once the cluster state is prepared, the
computation proceeds deterministically, i.e. \emph{every}
measurement produces a meaningful result, requiring only single
qubit measurements and feed-forward of the measurement result.
Feed-forward is the essential feature that makes one-way quantum
computing deterministic and can be seen as an active correction
of errors introduced by the randomness of measurement outcomes. We
will argue in the following that present state-of-the-art
technology allows for a demonstration of deterministic
one-way quantum computing by implementing this active feed-forward technique.\\

A cluster state is a network of entangled qubits and represents a
universal state for quantum computing. Universal means that any
quantum logic operation can be carried out on a sufficiently large
and appropriately structured cluster state. These states arise
when individual qubits are prepared in the superposition state
$|+\rangle=(|0\rangle+|1\rangle)/\sqrt{2}$, where
$|0\rangle,|1\rangle$ denote the computational basis states, and
connected by applying a controlled-PHASE operation
$|j\rangle|k\rangle\rightarrow(-1)^{jk}|j\rangle|k\rangle$ with
$(j,k\,\epsilon\,0,1)$ between neighboring qubits, effectively
generating entanglement. Recent experiments succeeded in creating
cluster states with various
methods~\cite{Walther05,Kiesel_2005,Zhang_2006}, including linear
optical realizations of simple controlled-PHASE
gates~\cite{Langford_2005,Kiesel_2005b,Okamoto_2005}.\\

Single qubit measurements are essential in cluster state quantum
computing. The shape of the cluster state and the nature of these
measurements, i.e. the order of measurements and the individual
measurement bases are determined by the desired algorithm. The
input state $|\psi_{in}\rangle$ is always initialized as
$|+\rangle$. It is important to note that the entire information
of the input state is initially stored in the multi-particle
correlations of the cluster, with the individual physical qubits
being completely undefined and therefore not carrying any
information about the input state. In this sense, namely that
properties of individual subsystems are completely undefined, the
cluster state is a maximally entangled state. Well-known examples
include 2-qubit Bell states and 3-qubit GHZ states. Single qubit
measurements on the cluster processes the encoded input from one
qubit to another analogous to remote state preparation. In
principle, two basic types of single-particle measurements
suffice to operate the one-way quantum computer. Measurements in
the computational basis $\{|0\rangle_{j},|1\rangle_{j}\}$ have the
effect of disentangling, i.e., removing the physical qubit $j$
from the cluster. This leaves a smaller cluster state and thus
gives the ability to shape the cluster to the specific algorithm.
The measurements which perform the actual quantum information
processing are made in the basis
$B(\alpha)=\{|\alpha_{+}\rangle,|\alpha_{-}\rangle\}$, where
$|\alpha_{\pm}\rangle=(|0\rangle\pm e^{
-i\alpha}|1\rangle)/\sqrt{2}$ with $\alpha\epsilon[0,2\pi]$. For
simplicity, we will restrict our discussion on single-qubit gate
operations, i.e. measurements on linear cluster
states~\cite{Walther05}. The argument can be generalized in a
straight-forward manner. \\
The choice of measurement basis determines the single-qubit
rotation, $R_{z}(\alpha)=\exp(-i\alpha\sigma_{z}/2)$ , followed
by a Hadamard operation, $H=(\sigma_{x}+\sigma_{z})/\sqrt{2}$, on
the input state ($\sigma_{x}$, $\sigma_{y}$, $\sigma_{z}$, being
the Pauli matrices).
\begin{equation}\label{gate}
R_{z}(\alpha)H|\psi_{in}\rangle\quad\Rightarrow\quad|\psi_{in}\rangle\
\Qcircuit @C=1em @R=.7em {
& \qw & \gate{R_{z}^{ (\alpha)}} & \gate{H} & \qw &\ \ |\psi_{out}\rangle\\
} \end{equation} The order and choices of these measurements
determine the unitary gates that are implemented and therefore the
algorithm that is computed. Remember that input states are by
construction always $|\psi_{in}\rangle=|+\rangle$ unless the
cluster is part of a larger cluster state. Rotations around the
z-axis can be implemented through the identity
$HR_{z}(\alpha)H=R_{x}(\alpha)$ so that two consecutive
measurements on a linear 3-qubit cluster can rotate the input
state to any arbitrary output state on the Poincare-Sphere
\begin{equation}\label{comp}
    R_{z}(\alpha)HR_{z}(\beta)H|\psi_{in}\rangle=R_{z}(\alpha)R_{x}(\beta)|\psi_{in}\rangle\quad\Rightarrow\quad|\psi_{in}\rangle\
\Qcircuit @C=1em @R=.7em {
& \qw & \gate{R_{z}^{ (\alpha)}} & \gate{R_{x}^{(\beta)}} & \qw &\ \ |\psi_{out}\rangle\\
} .
\end{equation}
Up until now, we have not incorporated the actual measurement
result in our analysis. Eq.~\ref{gate} only holds if the outcome
of the measurement $s$ is as desired, say $s=0$. Due to the
intrinsic randomness of the quantum measurement, it happens with
equal probability that the measurement yields the unwanted result
$s=1$. In that case, a well known Pauli-error
($\sigma_{x}=\Qcircuit @C=1em @R=.7em {
& \gate{X} & \qw\\
}$) is introduced in the computation, so that the single
measurement in basis $B_{j}(\alpha)$ rotates the qubit to:
\begin{equation}\label{comp1}
R_{z}(\beta)H\sigma_{x}|\psi_{in}\rangle\quad\Rightarrow\quad|\psi_{in}\rangle\
\Qcircuit @C=1em @R=.7em {
& \qw & \gate{R_{z}^{ (\alpha)}} & \gate{H} & \gate{X} & \qw &\ \ |\psi_{out}\rangle\\\
}
\end{equation}
Obviously, by adapting the measurement bases of subsequent
measurements, these errors can be eliminated. In the following,
let us consider the general case of a single-qubit operation by
taking into account the feed-forward rules. If we choose
consecutive measurements in bases $B_{1}(\alpha)$ and
$B_{2}(\beta)$ on physical qubits 1 and 2 of a 3-qubit cluster,
then we rotate the encoded input qubit $|\psi_{in}\rangle$ to the
output state
\begin{equation}\label{clusterstate}
|\psi_{out}\rangle=\sigma_{x}^{s_{2}}HR_{z}((-1)^{s_{1}}\beta)\sigma_{x}^{s_{1}}HR_{z}(\alpha)|\psi_{in}\rangle=\sigma_{x}^{s_{2}}\sigma_{z}^{s_{1}}R_{x}((-1)^{s_{1}}\beta)R_{z}(\alpha)|\psi_{in}\rangle
\end{equation} which is stored on qubit 3. The measurement outcome,
$s_{i}=\{0,1\}$, on the physical qubit $i$ determines the
measurement basis for the succeeding qubit and indicates any
introduced Pauli errors that have to be compensated for. This
idea can schematically be depicted as a circuit diagram:
\[
\Qcircuit @C=1.8em @R=1.2em  {
\lstick{\ket{\psi_{in}}}   & \ctrl{1}&  \measure{\mbox{$B_{1}(\alpha)$}} & \meter \cwx[1]& \cw & \control\cw\cwx[1]\\
\lstick{\ket{+}}  & \ctrl{1} &  \qw& \measure{\mbox{$B_{2}(\pm\beta)$}} & \meter & \control \cw\\
\lstick{\ket{+}}  & \ctrl{-1}  & \qw & \qw & \gate{X} \cwx &
\gate{Z} \cwx & \rstick{\ket{\psi_{out}}} \qw \\
&  Cluster & & & \qquad\quad\ Error\ Correction
\gategroup{1}{2}{3}{2}{1.5em}{.}
\gategroup{3}{5}{3}{6}{0.5em}{--} }
\]
Single wires represent quantum channels, while double lines denote
classical communication. The circles in front of the measurement
meters show the measurement basis. No error correction is required
for the specific case where the outcomes of the first and second
qubit are $s_{1}=s_{2}=0$ and hence, as expected,
$|\psi_{out}\rangle=R_{x}(\beta)R_{z}(\alpha)|\psi_{in}\rangle$.
However, if the outcome of the second qubit is $s_{1}=1$
($s_{2}=0$) the measurement basis of the third qubit has to be
changed from $B_{2}(\beta)$ to $B_{2}(-\beta)$ and finalized by a
Pauli error correction, i.e. $\sigma_{z}$ on the output qubit, to
get the desired output of the computation. This yields
$|\psi_{out}\rangle=\sigma_{z}R_{x}(-\beta)R_{z}(\alpha)|\psi_{in}\rangle$
Similar corrections are required in the cases when the third
qubit's outcome is $s_{2}=1$ ($s_{1}=0$) and hence
$|\psi_{out}\rangle=\sigma_{z}R_{x}(\beta)R_{z}(\alpha)|\psi_{in}\rangle$.
Finally, if an unwanted projection occurs to both qubits,
($s_{1}=s_{2}=1$), two Pauli errors, $\sigma_{z}$ and
$\sigma_{x}$, have to be compensated for on qubit 3 yielding
$|\psi_{out}\rangle=\sigma_{x}\sigma_{z}R_{x}(-\beta)R_{z}(\alpha)|\psi_{in}\rangle$.
This is summarized in Table 1.

Experimentally, feed-forward can only be achieved by recording
both measurement outcomes simultaneously, $s_{i}=\{0,1\}$. The
recent photonic realization of a one-way quantum
computer~\cite{Walther05} employed single-port polarizers, which
are, although sufficient to demonstrate the working principle,
not suited for this purpose. Simultaneous recording of the
measurement results can be achieved with polarizing beamsplitters
(PBSs), preceded by half- and quarter-wave plates to chose
arbitrary measurement bases. The basis of the measurements can be
adapted by employing fast-switching and low-loss electro-optical
modulators (EOMs), which, depending on the applied voltage,
change the photon's state of polarization. Analogously,
error-correction can be performed on the output qubit if the EOMs
are aligned to apply $\sigma_{x}$ and $\sigma_{z}$ rotations,
respectively.

In an experimental implementation of this scheme, the individual
photonic qubits must be delayed just long enough so that the
classical feed-forward process can be carried out, i.e., that an
individual outcome can adapt the measurement basis for the next
measurement. The most rudimentary "quantum memory" that can be
used for such purpose is a single-mode fiber of a specific length,
which has negligible photon loss over moderate distances. Every
single feed-forward process includes detection of the photon,
processing of the measurement result and finally switching of the
modulator to adapt the measurement basis in real time and/or
performing error correction on the output qubit. A major advantage
of optical quantum computation is the achievable high speed of the
gate operation. Various types of EOMs achieve low-loss and high
contrast switching with fidelities above 99\%. Switching times are
well below 100~ns when combined with custom built drivers and such
devices have successfully been implemented in early
demonstrations of feed-forward
control~\cite{Ursin04,Giacomini02,Pittman_2002}. Currently
available logic boards and single-photon detectors have response
times of around 10~ns and 30~ns, respectively, so that
feed-forward cycles of less than 150~ns seem experimentally
feasible. This time-scale corresponds to a single-mode fiber delay
line of approximately 30~m. A gate time of 150-300~ns for one
computational step is, to our best knowledge, about three orders
of magnitude faster than achievable in other physical
realizations of quantum computers such as in
ion-traps~\cite{Riebe04,Barrett04} or in NMR~\cite{Lieven_2001}.\\

Based on our recent successful demonstration of one-way quantum
computing~\cite{Walther05}, a proof-of-concept demonstration of
\emph{deterministic }quantum computing, i.e. implementation of
active feed-forward and error-correction in real time, on a
4-photon cluster state is certainly feasible. Conceptually, this
would present a crucial step towards realizing scalable optical
quantum computing, showing that it is indeed possible to build a
deterministic quantum computer which uses both entanglement and
the intrinsically random measurement outcomes as an essential
feature.

\section{Entanglement as communication channel -- Quantum Communication}
\subsection{Distributed Computing: Entanglement for Quantum Communication
Complexity}
Although entanglement on its own cannot be used for
communication, it surprisingly {\it can produce effects as if
information had been transferred}. In a communication complexity
problem, separated parties performing {\it local} computations
exchange information in order to accomplish a {\it globally}
defined task, which is impossible to solve
single-handedly~\cite{Yao1,Yao2}. Remarkably, if the parties share
entanglement the required information exchange in the
communication complexity problem can be reduced~\cite{Cleve1997}
or even eliminated~\cite{Buhrman1997}. Such a reduction of
communication complexity might be important in future for
speeding up distributed computations, e.g.\ within very large
scale integration (VLSI) circuits.

Here we will determine the experimental requirements for quantum
communication complexity protocols to outperform their classical
counterparts in solving certain types of problems. This will
include determination of the required minimal visibility $V$ and
the detection efficiency $\eta$ for the advantage. The type of the
problems considered here is as follows. There are $n$ separated
partners who receive local input data $x_i$ such that they know
only their own data and not those of the partners. The goal is for
all of them to determine the value of a function $f(x_1,...,x_n)$.
Before they start the protocol, they are allowed to share
classically correlated random strings or quantum entanglement. If
only a {\it restricted} amount of communication is allowed, we ask
the questions: {\it What is the highest possible probability for
the parties to arrive at the correct value of the function?} We
refer to this probability as "success rate" of the protocol.

Recently, it has been realized that communication complexity
problems are tightly linked to Bell's theorem~\cite{Bell64}. On
the basis of this insight, quantum protocols are developed that
exploit entanglement between qubits~\cite{Brukner2004},
qutrits~\cite{Brukner02} and higher dimensional
states~\cite{Brukner2003}. The crucial idea is that every
classical protocol can be simulated by a local realistic model
and thus its success rate is limited by the Bell-type
inequalities~\cite{Brukner2004}. In contrast, the success rate of
quantum protocols---which make use of entangled states---can
exceed these limits, since entangled states are at variance with
local realism. More precisely, for every Bell's inequality---even
those which are not yet known---there exists a communication
complexity problem, for which a protocol assisted by states which
violate the inequality has a higher success rate than any
classical protocol. Violation of Bell's inequalities is thus the
{\it necessary and sufficient condition} for quantum protocols to
beat the classical ones.

Consider the general Bell's inequality for correlation functions
\begin{equation}
\sum_{x_1,...,x_n=0}^{1} g(x_1,...,x_n) E({x_1},...,{x_n}) \leq
B(n). \label{vatra}
\end{equation}
Here $g$ is a real function, $B(n)$ is a bound imposed by local
realism and $E({x_1},...,{x_n})$ is the correlation function for
measurements on $n$ particles, which involve, at each local
measurement station $i$, two alternative dichotomic observables,
parameterized here by $x_i\!=\!0$ and $1$. In
Ref.~\cite{Brukner2004} it was shown that this Bell's inequality
puts limits on the success rate in computation of certain
two-valued functions $f(x_1,...,x_n)$ with the inputs $x_i=0$ or
$1$~\cite{comment}. The execution of the protocol is successful
when {\em all} parties arrive at the correct value of $f$.

The most interesting case found is  for $ g\!=\!\sqrt{2^{n+1}}
\cos\left[\frac{\pi}{2}(x_1 \! + \!...\!+\! x_n)\right]$, $n$ odd
and $B(n) \!=\! 2^n$ for which the success probability of
classical solutions cannot be larger than
\begin{equation}
P_{\text{class}}\!=\!\frac{1}{2} \left(1\!+\!
\frac{1}{\sqrt{2^{n-1}}}\right)\!,
\end{equation}
whereas a quantum protocol solves the problem with certainty,
i.e.\ $P_{\text{quant}}=1$~\cite{comment}. This implies that in
the limit of very large $n$ one has $P_{\text{class}}\!
\rightarrow \!1/2$, which is not better than if the partners
simply agree beforehand to choose all the same (random) value for
the value of the function.

Without going into the details of the protocols we mention here
that both in the classical and quantum case the partners give all
the {\it same} value for their guess of the value of the function
$f$. (This value is obtained as a product of $n$ locally produced
values $e_i$, where $e_i$ is broadcasted by party $i$.
See~\cite{Brukner2004,comment} for details.) The important
difference is that in a quantum protocol this value is obtained
from local results of the Bell experiment for $n$ parties,
whereas in a classical protocol it is obtained from the results
of local (classical) operations assisted with classical
correlations. The maximal success rate of $P_{\text{quant}}=1$ of
the quantum protocol is obtained using the
Greenberger-Horne-Zeilinger state $|GHZ\rangle =
(|0\rangle_1...|0\rangle_n+|1\rangle_1...|1\rangle_n)/\sqrt{2}$~\cite{GHZ89}.

For the quantum protocol to beat the best classical one we need a
success higher than $P_{\text{class}}$. We now analyze detectors
with finite detection efficiency $\eta$ and non-maximal
visibility $V$ due to experimental imperfections as modeled by an
admixture of white noise to the perfect state: $\rho = V
|GHZ\rangle \langle GHZ|+(1-V)\, I/2^n$.


With a finite detector efficiency $\eta \leq 1$, the partners
obtain perfect quantum correlations in $\eta^n V$ of the cases and
proceed with the quantum protocol with the success rate
$P_{\text{quant}}=1$. When their detectors fail, the partners must
agree on a procedure. They are not allowed to communicate the
failure, as this would consist of further bits of communication
between the parties, and the allowed communication is restricted.
The most effective way for a partner is to proceed with the best
classical protocol in case her/his detector fails. It is assumed
that there are no experimental constraints for classical protocols
as they are based on manipulating and detecting classical systems
(e.g.\ balls or pencils), which could be done with very high
efficiency.

Whenever all detectors fail, which happens in $(1-\eta)^n$ of the
cases, the partners will obtain the best classical success rate
$P_{\text{class}}$. In the cases when some of the detectors fail
and the rest fire, the partners whose detectors fail would start
the best classical protocols, whereas those whose detectors fire
proceed with the quantum protocol. Since the two protocols are
completely independent, the success rate is not better than the
probability that all partners give the same but random guess for
the value of the function. In the rest of the cases, all detectors
fire measuring white noise, which again leads to the success as
for the random guess. Thus, in $1-\eta^n V-(1-\eta)^n$ of the
cases the success rate is $P_{\text{rand}}=1/2$.

Taking all this into account, the condition for a
higher-than-classical success rate is:
\begin{equation}
\eta^n V + (1-\eta)^n P_{\text{class}} + (1-\eta^n V-(1-\eta)^n)
\,\tfrac{1}{2} > P_{\text{class}}. \label{inequality}
\end{equation}
A similar analysis for the special case of $n=3$ and special
function $f$ was given by Galvao in Ref.~\cite{Galvao2002}. In
Figure 1 we show the region in the parameter space of $V$, $\eta$
and $n$ that guarantees a higher-than-classical success rate.
Taking $\eta=0.8$ for the detector efficiency and visibility
$V=0.9$, one obtains $n=4$ for the minimal number of photons in
the entangled state, which is well within the scope of current
technology. Recently, a quantum communication complexity protocol
based on the sequential transfer of a {\it single}
qubit~\cite{Galvao2002} was experimentally implemented and its
advantage over the classical counterpart was shown in the
presence of the imperfections of a state-of-the-art
set-up~\cite{Trojek2005}. It could therefore be expected in near
future that entanglement-based quantum communication complexity
protocols will become comparable to quantum key distribution, the
only commercial application of quantum information science so far.

\subsection{Distributed Entanglement in Higher Dimension: Entangled Qutrit Quantum Cryptography}
All Quantum Cryptography experiments performed so far were based
on two-dimensional quantum systems (qubits). However, the usage of
higher-dimensional systems offers advantages such as an increased
level of tolerance to noise at a given level of security and a
higher flux of information compared to the qubit cryptography
schemes.\\

In a recent experiment we produced two identical keys using, for
the first time, entangled trinary quantum systems (qutrits) for
quantum key distribution~\cite{Groeblacher2006}. The advantage of
qutrits over the normally used binary quantum systems is an
increased coding density and a higher security margin of 22\%
(instead of ca. 15\%). The qutrits are encoded into the orbital
angular momentum of photons, namely Laguerre-Gaussian modes with
azimuthal index $l +1, 0$ and $-1$, respectively. The orbital
angular momentum is controlled with static phase holograms. In an
Ekert-type protocol the violation of a three-dimensional Bell
inequality verifies the security of the generated keys. A key is
obtained with a qutrit error rate of approximately 10\%.  The
security of this key is ascertained by the violation of the Bell
inequality, with $S = 2.688 \pm 0.171$. In contrast to the
polarization degree of freedom, in principle there is no
limitation on the dimension of the two-photon entanglement with
orbital angular momentum and therefore an extension of the qutrit
to a more general qudit case is feasible. This opens up a new
class of experiments with higher dimensional entanglement.\\

Spatial light modulators (SLM) promise a fascinating new
experimental approach for working with the orbital angular
momentum of photons. The main idea is to use the SLM for applying
computer calculated holograms to the entangled photons (see
Fig.~\ref{trits}) instead of static phase plates. Thereby we gain
huge experimental flexibility, since we are now able to
superimpose several optical elements such as lens configurations,
mirrors and phase singularity onto one active phase element, and
fine tune the holograms simply by adjusting the parameters in the
calculation. This will open up possibilities of further study of
three dimensional entanglement (or more dimensions), which is an
area with many unknown features. A first successful demonstration
of this method is shown in Figure~\ref{trits} where we analyzed
the correlation of entangled photons, where the orbital angular
momentum of one of the photons is transformed via the SLM.

\subsection{Distributed Entanglement: Long Distance Quantum
Communication \& Quantum Networking} There is a range of unique
applications emerging if several users share entangled particles,
such as quantum
cryptography~\cite{Bennett84,Ekert91,Jennewein00b,Naik00,Tittel00},
quantum teleportation~\cite{Bennett93,Bouwmeester97}, quantum
dense coding~\cite{Bennett92b,Mattle96} or communication
complexity (see previous section). Clearly it is an important
prerequisite to be able to establish networks of quantum
communication, similar to what classical communication networks
do. It is particularly desirable to establish entanglement
between several users, with a very flexible network hierarchy.
For example, two users who wish to share entangled particles just
call their network operator, who performs the necessary settings
to accomplish this task. Likewise, if three users wish to share
GHZ-states, again the network operator performs the required
operations for this task. \\
Fortunately, quantum physics allows us to perform these tasks, if
the several users initially share entangled particles with a
central network operator. Utilizing the procedure known as
entanglement swapping~\cite{Zukowski95,Jennewein02}, the
generalization of quantum teleportation, the operator may simply
swap the entanglement between the particles entangled with two
different users, such that finally the particles of the two users
get entangled. The operations that the central node (operator)
must perform are projection measurements onto the desired
entangled state. Since the particles originally have no relation,
the projective measurement will give a random result, which must
be communicated to the users, so they can use the entangled
particles. Entanglement swapping can in principle be generalized
to arbitrary quantum network sizes if the network operator
performs the swapping operations (i.e. projections on to
Bell-states, GHZ-states), depending on which users wish to
communicate. This is at the heart of a quantum
repeater~\cite{Briegel98}, which additionally makes use of
entanglement purification~\cite{Bennett96,Pan03b} and quantum
memories to faithfully transmit entanglement over arbitrary
distances. Important experimental progress has been made along
this line, for example by demonstrating quantum teleportation
over long distances~\cite{Ursin04} or by realizing non-classical
interference of photons from completely independent photon
sources~\cite{Kaltenbaek2006}.\\
In the future, the use of satellite-based technology could
provide the means for distribution of quantum signals even on a
global scale~\cite{Nordholt02,Rarity02,Aspelmeyer03b}. These
schemes will involve sources for entangled photons onboard
satellites, which are sent via telescopes to other satellites as
well as optical earth-based ground stations. The principles of
this concept, free-space quantum communication, have been
demonstrated in various experiments both for faint-pulse
systems~´\cite{Hughes02,Rarity_2001,kurtsiefer02} and for
entangled photons~\cite{Aspelmeyer03,Resch05,Peng05,Ursin_2006} .
The current distance record has been only recently achieved in a
144~km inter-island link using entangled
photons~\cite{Ursin_2006}. These result are very promising for
entanglement-based free-space quantum communication in
high-density urban areas and even for large distances. It is also
encouraging for optical quantum communication between ground
stations and satellites since the length of our free-space link
exceeds the atmospheric
equivalent.\\
The progress of this huge research program of extending quantum
communication to Space is advancing and is taking on more momentum
continuously. In mid 2005 we have received the official positive
response to our proposal to the ELIPS-2 announcement for
opportunity, by the European Space Agency (ESA), describing our
experiment called "Quantum entanglement for Space experiments",
Space-QUEST, in full detail. The clear aim of our proposed
experiment is to place an entangled photon source on-board the
ISS and send the two photons towards two receiving ground
stations. This will allow performing fundamental experiments,
since the entangled photons can be separated by up to 1500~km
distance hence providing a significant enlargement of the size of
a quantum state possible on ground. Furthermore, the experiment
allows demonstrations of quantum communication applications at a
global scale, which is clearly not feasible with ground based
systems. Presently we are working together with scientific and
industrial partners on further refinements of the planned
systems, as well as with funding agencies to reserve the
resources for this large-scale experiment. In particular, we have
performed a design study together with Contraves AG (CH) for a
quantum communication terminal based on existing laser
communication terminals, to be placed on the International Space
Station \cite{Pfennigbauer05}. The designed platform contains all
optical, laser and electronic components required for the quantum
terminal operation. This system is based on the OPTEL25 optical
terminal, from Contraves AG (CH), designed for inter-satellite
laser communication. In addition we are performing
proof-of-concept experiments over 144 km using an inter-island
link between La Palma and Teneriffe.

A very interesting approach, alternative to the ISS system, is to
implement quantum communication uplinks from ground to
satellites. This scheme is particularly interesting, as the
technical complexity of the Space-based receiver is significantly
simpler than for the full quantum communication transmitter.
Thereby the technical difficulty is transferred to the ground
segment, which is clearly a well available environment. The main
technical difficulty is the implementation of an adaptive optics
system capable of pre-compensating the wave front distortion that
the uplink beam will experience as it traverses the atmosphere.


If only one receiver in Space is implemented, this scheme allows
quantum communication, such as renewing keys for the
communication with satellites via quantum cryptography, or the
key exchange between separate ground stations, by joining up the
successive keys. If two receivers are realized, they would allow
performing also fundamental tests of quantum entanglement over
huge distances, and also with high relative velocities between
the observers~\cite{Kaltenbaek04a}. When the receiver satellite is
realized as a double-receiver unit, it will allow to combine
uplinks a) and d) of Fig.~\ref{uplinks} and perform quantum
entanglement swapping for these photons, which will be an
important step towards global quantum communication networks by
distributing entanglement. The most critical technology is the
required adaptive optics for reducing the "shower curtain" effect
of the atmosphere. A ground-to-Space link suffers much stronger
from atmospheric induced beam deviation, since the errors are
induced at the beginning of the beam path. As a consequence, the
transmitter on the ground must (pre-)compensate the distortion of
the transmitted light beam in order to minimize the spotsize
received at the satellite.
\\
The advantage of this scenario is that the main technological
difficulties are transferred to the ground segment, whereas the
Space-segment is a somewhat simple receiver. The key
technological elements at the ground based transmitter stations
are (Fig.~\ref{uplink_sys}): (1) The source of entangled photons:
since placed on the ground, the limits on the size, power, and
stability of a source are highly relaxed. This would allow using
a down conversion source pumped by a high power solid laser (e.g.
Nd-YAG) laser, which already today produces high quality
entangled photons at a very high rate (5~M~pairs/s). (2) High
quality transmitter telescopes: there are several suitable
optical ground stations available with apertures between 0.5 --
1.5~m diameter and diffraction limited quality, which offer the
ability to track low flying satellites. (3) Adaptive optics: This
is the crucial technology required for achieving link distances
of several thousands of km. Thereby the active wavefront
transformation will precompensate the wavefront distortion that
the beam of photons will experience when traveling through the
atmosphere. This technology is still in a laboratory status, and
has yet to be applied to the specific needs of quantum
communication. Since this technology could be utilize on the
ground segment the technology can be realized at the highest
possible level available at the time of the experiments.
\\
The key technological elements at the satellite based receiver
are (Fig.~\ref{uplink_sys}): (1) Receiver telescope: the optical
quality of the receiving telescope must not be diffraction
limited, if the field of view (FOV) of the receiver is
sufficiently large. This is finally limited by the diameter of
the detectors. The telescopes must be motorized, and also employ
a fine-pointing mechanism for keeping up the alignment with the
transmitter telescope on ground. The tracking is performed with
the signals measured from reference lasers operating at a
separate wavelength. (2) Analysis of the photons: the polarization
of the photons must be analyzed in two different bases, such as
horizontal-vertical and diagonal linear polarization. This can be
accomplished with a combination of a symmetric beam splitter and
polarizing beam splitters in each output arm. In addition,
spectral filtering of the photons must be utilized in order to
suppress the background light. (3) Detection of the photons: The
received photons are focused on single photon detectors.
Depending on the final choice of wavelength, the presently best
single photon detection is achieved with silicon-avalanche photo
diodes (Si-APD). (4) Acquisition and processing of the detection
events: the detection events must be recorded as time-tags, and
stored on the satellite. The data must be sent to the ground for
further processing.


\section{Conclusion}

In summary, we have introduced and reviewed some recent
experimental progress in the understanding of photonic quantum
entanglement as a resource for quantum information processing. We
have also provided an outlook onto future experiments that should
be feasible with current technology and that will further
highlight the distinctive role of entanglement. Besides the
impressive achievements in laboratories all over the world there
remain fascinating challenges for the future ranging from the
interfacing of photons to scalable and durable architectures,
i.e. including quantum memories, over the faithful production and
characterization of multipartite entangled states of significant
particle number to the realization of a full scale quantum
repeater.

\begin{acknowledgements}
We thank J. Kofler for valuable discussions. We acknowledge
financial support by the FWF Austrian Science Fund, SFB project
P06, the European Commission under the Integrated Project Qubit
Applications QAP funded by the IST Directorate as contract number
015846, the ARO-funded U.S. Army Research Office Contract No.
W911NF-05-0397 and the City of Vienna.
\end{acknowledgements}

\newpage

\begin{thebibliography}{10}
\newcommand{\enquote}[1]{``#1''}
\expandafter\ifx\csname url\endcsname\relax
  \def\url#1{\texttt{#1}}\fi
\expandafter\ifx\csname
urlprefix\endcsname\relax\def\urlprefix{URL }\fi
\providecommand{\eprint}[2][]{\url{#2}}

\bibitem{Schrodinger35a}
E.~Schr{\"o}dinger, \enquote{Die Gegenw{\"a}rtige {S}ituation in
der
  {Q}uantenmechanik,} Naturwissenschaften \textbf{23}, 807--812; 823--828;
  844--849 (1935).

\bibitem{Wu_1950}
C.~S. Wu and I.~Shaknov, \enquote{The angular correlation of
annihilation
  radiation,} Phys.\ Rev. \textbf{77}, 136 (1950).

\bibitem{Kocher_1967}
C.~A. Kocher and E.~D. Commins, \enquote{Polarization correlation
of phootns
  emitted in an atomic cascade,} Phys.\ Rev.\ Lett. \textbf{18}, 575--577
  (1967).

\bibitem{Clauser72}
J.~F. Clauser, \enquote{Experimental Limitations to the Validity
of
  Semiclassical Radiation Theories,} Phys.\ Rev.\ A \textbf{6}, 49--54 (1972).

\bibitem{Clauser74}
J.~F. Clauser, \enquote{Experimental distinction between the
classical and
  quantum field-theoreti predictions for the photoelectric effect,} Phys.\
  Rev.\ D \textbf{9}, 853--860 (1974).

\bibitem{Bell64}
J.~S. Bell, \enquote{On the {E}instein {P}odolsky {R}osen
Paradox,} Physics
  \textbf{1}, 195--200 (1964).

\bibitem{Freedman72}
S.~J. Freedman and J.~F. Clauser, \enquote{Experimental Test of
Local
  Hidden-Variable Theories,} Phys.\ Rev.\ Lett. \textbf{28}, 938--941 (1972).

\bibitem{Aspect82b}
A.~Aspect, J.~Dalibard, and G.~Roger, \enquote{Experimental Test
of Bell's
  Inequalities Using Time- Varying Analyzers,} Phys.\ Rev.\ Lett. \textbf{49},
  1804--1807 (1982).

\bibitem{Weihs98}
G.~Weihs, T.~Jennewein, C.~Simon, H.~Weinfurter, and A.~Zeilinger,
  \enquote{Violation of {B}ell's Inequality under Strict {E}instein Locality
  Conditions,} Phys.\ Rev.\ Lett. \textbf{81}, 5039--5043 (1998).

\bibitem{Wiesner83}
S.~Wiesner, \enquote{Conjugate coding,} Sigact News \textbf{15},
78--88 (1983).

\bibitem{Bennett84}
C.~H. Bennett and G.~Brassard, \enquote{Quantum cryptography:
Public Key
  Distribution and coin-tossing,} Proceedings of IEEE International Conference
  on Computers, Systems and Signa Processing, Bangalore, India pp. 175--179
  (1984).

\bibitem{Ekert91}
A.~K. Ekert, \enquote{Quantum Cryptography Based on Bell's
Theorem,} Phys.\
  Rev.\ Lett. \textbf{67}, 661--663 (1991).

\bibitem{Bennett93}
C.~H. Bennett, G.~Brassard, C.~Cr{\'e}peau, R.~Jozsa, and A.~P.
  an~W.~K.~Wootters, \enquote{Teleporting an unknown quantum state via dual
  classical and Einstein-Podolsky-Rose channels,} Phys.\ Rev.\ Lett.
  \textbf{70}, 1895--1899 (1993).

\bibitem{Bouwmeester97}
D.~Bouwmeester, J.-W. Pan, K.~Mattle, M.~Eibl, H.~Weinfurte, and
A.~Zeilinger,
  \enquote{Experimental Quantum Teleportation,} Nature \textbf{390}, 575--579
  (1997).

\bibitem{Bennett92b}
C.~H. Bennett and S.~J. Wiesner, \enquote{Communication via One-
and
  Two-Particle Operators on {E}instein-{P}odolsky-{R}ose States,} Phys.\ Rev.\
  Lett. \textbf{69}, 2881--2884 (1992).

\bibitem{Mattle96}
K.~Mattle, H.~Weinfurter, P.~G. Kwiat, and A.~Zeilinger,
\enquote{Dense Coding
  in Experimental Quantum Communication,} Phys.\ Rev.\ Lett. \textbf{76},
  4656--4659 (1996).

\bibitem{Knill00}
E.~Knill, R.~Laflamme, and G.~Milburn, \enquote{A Scheme for
Efficient Quantum
  Computation with Linear Optics,} Nature \textbf{409}, 46--52 (2000).

\bibitem{Kok_2005}
P.~Kok, W.~Munro, K.~Nemoto, T.~Ralph, J.~P. Dowling, and
G.~Milburn,
  \enquote{Review article: Linear optical quantum computing,} quant-ph/0512071
  (2005).

\bibitem{Raussendorf01}
R.~Raussendorf and H.~J. Briegel, \enquote{A One-Way Quantum
Computer,} Phys.
  Rev. Lett. \textbf{86}(22), 5188--5191 (2001).

\bibitem{Nielsen04b}
M.~A. Nielsen, \enquote{Optical quantum computation using cluster
states,}
  Phys.\ Rev.\ Lett. \textbf{93}, 040,503 (2004).

\bibitem{Walther05}
P.~Walther, K.~J. Resch, T.~Rudolph, E.~Schenck, H.~Weinfurter,
V.~Vedral,
  M.~Aspelmeyer, and A.~Zeilinger, \enquote{Experimental one-way quantum
  computing,} Nature \textbf{434}, 169--176 (2005).

\bibitem{Kiesel_2005}
N.~Kiesel, C.~Schmid, U.~Weber, G.~Toth, O.~G\"uhne, R.~Ursin, and
  H.~Weinfurter, \enquote{Experimental Analysis of a Four-Qubit Photon Cluster
  State,} Phys.\ Rev.\ Lett. \textbf{95}, 210,502 (2005).

\bibitem{Zhang_2006}
A.-N. Zhang, C.-Y. Lu, X.-Q. Zhou, Y.-A. Chen, Z.~Zhao, T.~Yang,
and J.-W. Pan,
  \enquote{Experimental construction of optical multiqubit cluster states from
  Bell states,} Phys.\ Rev.\ A \textbf{73}, 022,330 (2006).

\bibitem{Langford_2005}
N.~K. Langford, T.~J. Weinhold, R.~Prevedel, K.~J. Resch,
A.~Gilchrist, J.~L.
  O'Brien, G.~J. Pryde, and A.~G. White, \enquote{Demonstration of a Simple
  Entangling Optical Gate and Its Use in Bell-State Analysis,} Phys.\ Rev.\
  Lett. \textbf{95}, 210,504 (2005).

\bibitem{Kiesel_2005b}
N.~Kiesel, C.~Schmid, U.~Weber, R.~Ursin, and H.~Weinfurter,
\enquote{Linear
  Optics Controlled-Phase Gate Made Simple,} Phys.\ Rev.\ Lett. \textbf{95},
  210,505 (2005).

\bibitem{Okamoto_2005}
R.~Okamoto, H.~F. Hofmann, S.~Takeuchi, and K.~Sasaki,
\enquote{Demonstration
  of an Optical Quantum Controlled-NOT Gate without Path Interference,} Phys.\
  Rev.\ Lett. \textbf{95}, 210,506 (2005).

\bibitem{Ursin04}
R.~Ursin, T.~Jennewein, M.~Aspelmeyer, R.~Kaltenbaek,
M.~Lindentha, P.~Walther,
  and A.~Zeilinger, \enquote{Quantum teleportation across the Danube,} Nature
  \textbf{430}, 849 (2004).

\bibitem{Giacomini02}
S.~Giacomini, F.~Sciarrino, E.~Lombardi, and F.~D. Martini,
\enquote{Active
  teleportation of a quantum bit,} Phys.\ Rev.\ A \textbf{66}, 030,302(R)
  (2002).

\bibitem{Pittman_2002}
T.~B. Pittman, B.~C. Jacobs, and J.~D. Franson,
\enquote{Demonstration of
  feed-forward control for linear optics quantum computation,} Phys.\ Rev.\ A
  \textbf{66}, 052,305 (2002).

\bibitem{Riebe04}
M.~Riebe, H.~H{\"a}ffner, C.~F. Roos, W.~H{\"a}nsel, J.~B.
  an~G.~P.T.~Lancaster, T.~W. K{\"o}rber, C.~Becher, F.~S.-K. an~D.~F.
  V.~James, and R.~Blatt, \enquote{Quantum teleportation with atoms,} Nature
  \textbf{429}, 734--737 (2004).

\bibitem{Barrett04}
M.~D. Barrett, J.~Chiaverini, T.~Schaetz, J.~Britton, W.~M.
Itano, J.~D. Jost,
  E.~Knill, C.~Langer, D.~Leibfried, R.~Ozeri, and D.~J. Wineland,
  \enquote{Quantum teleportation with atomic qubits,} Nature \textbf{429},
  737--739 (2004).

\bibitem{Lieven_2001}
L.~M. Vandersypen, M.~Steffen, G.~Breyta, C.~S. Yannoni, M.~H.
Sherwood, and
  I.~L. Chuang, \enquote{Experimental realization of Shor's quantum factoring
  algorithm using nuclear magnetic resonance,} Nature \textbf{414}, 883--887
  (2001).

\bibitem{Yao1}
A.~C.-C. Yao, in \emph{Proceedings of the 11th Annual ACM
Symposium on Theory
  of Computing}, pp. 209--213 (1979).

\bibitem{Yao2}
A.~C.-C. Yao, in \emph{Proceedings of the 34th Annual IEEE
Symposium in
  Foundations of Computer Science}, pp. 352--361 (1993).

\bibitem{Cleve1997}
R.~Cleve and H.~Buhrman, Phys.\ Rev.\ A \textbf{56}, 1201--1204
(1997).

\bibitem{Buhrman1997}
H.~Buhrman, R.~Cleve, and W.~van Dam, e-print quant-ph/9705033
(1997).

\bibitem{Brukner2004}
C.~Brukner, M.~Zukowski, J.-W. Pan, and A.~Zeilinger, Phys.\
Rev.\ Lett.
  \textbf{92}, 127,201 (2004).

\bibitem{Brukner02}
C.~Brukner, M.~Zukowski, and A.~Zeilinger, \enquote{Quantum
communication
  complexity protocol with two entangled qutrits,} Phys.\ Rev.\ Lett.
  \textbf{89}, 197,901 (2002).

\bibitem{Brukner2003}
C.~Brukner, T.~Paterek, and M.~Zukowski, Int. J. Quant. Inf.
\textbf{1},
  519--525 (2003).

\bibitem{comment}
Strictly speaking, in communication complexity problems of
  Ref.~\cite{Brukner2004} each party $i$ receives {\it two} one-bit inputs
  $(x_i,y_i)$ and their goal is to compute a function of the form
  $F(x_1,y_1,...,x_n,y_n)=y_1 \cdot ... \cdot y_n \cdot f(x_1,...,x_n)$. The
  values of the function $f$ and the $y_i$ are $\pm 1$. Each party is allowed
  to broadcast only {\it one bit} of information (denoted as $e_i$). For the
  present analysis the existence of inputs $y_i$ is not of importance and is
  ommited here. See Ref.~\cite{Brukner2004} for details.

\bibitem{GHZ89}
D.~M. Greenberger, M.~A. Horne, and A.~Zeilinger, \enquote{Going
beyond
  {B}ell's Theorem,} in \emph{Bell's Theorem, Quantum Theory, and Conceptions
  of the Universe}, M.~Kafatos, ed., p.~69 (Kluwer, Dordrecht, 1989).

\bibitem{Galvao2002}
E.~F. Galvao, \enquote{Feasible quantum communication complexity
protocol,}
  Phys.\ Rev.\ A \textbf{65}, 12,318 (2002).

\bibitem{Trojek2005}
P.~Trojek, C.~Schmid, M.~Bourennane, {\v C}.~Brukner,
M.~Zukowski, and
  H.~Weinfurter, \enquote{Experimental quantum communication complexity,}
  Phys.\ Rev.\ A \textbf{72}, 50,305(R) (2005).

\bibitem{Groeblacher2006}
S.~Gr\"oblacher, T.~Jennewein, A.~Vaziri, G.~Weihs, and
A.~Zeilinger,
  \enquote{Experimental quantum cryptography with qutrits,} N.\ J.\ Phys.
  \textbf{8}, 75 (2006).

\bibitem{Jennewein00b}
T.~Jennewein, C.~Simon, G.~Weihs, H.~Weinfurter, and A.~Zeilinger,
  \enquote{Quantum Cryptography with Entangled Photons,} Phys.\ Rev.\ Lett.
  \textbf{84}, 4729--4732 (2000).

\bibitem{Naik00}
D.~S. Naik, C.~G. Peterson, A.~G. White, A.~J. Berglund, and
P.~G. Kwiat,
  \enquote{Entangled State Quantum Cryptography: Eavesdropping on the Ekert
  Protocol,} Phys.\ Rev.\ Lett. \textbf{84}, 4733--4736 (2000).

\bibitem{Tittel00}
W.~Tittel, J.~Brendel, H.~Zbinden, and N.~Gisin, \enquote{Quantum
Cryptography
  Using Entangled Photons in Energy-Time Bell States,} Phys.\ Rev.\ Lett.
  \textbf{84}, 4737--4740 (2000).

\bibitem{Zukowski95}
M.~Zukowski, A.~Zeilinger, and H.~Weinfurter, \enquote{Entangling
Photons
  Radiated by Independent Pulsed Sources,} in \emph{Fundamental Problems in
  Quantum Theory: A Conference Held in Honor of Professo John A. Wheeler},
  D.~M. Greenberger and A.~Zeilinger, eds., Annals of the New York Academy of
  Sciences, pp. 91--102 (New York Academy of Sciences, New York, 1995).

\bibitem{Jennewein02}
T.~Jennewein, G.~Weihs, J.-W. Pan, and A.~Zeilinger,
\enquote{Experimental
  Nonlocality Proof of Quantum Teleportation and Entanglemen Swapping,} Phys.\
  Rev.\ Lett. \textbf{88}, 17,903 (2002).

\bibitem{Briegel98}
H.-J. Briegel, W.~D{\"u}r, J.~I. Cirac, and P.~Zoller,
\enquote{Quantum
  Repeaters: The Role of Imperfect Local Operations in Quantum Communication,}
  Phys.\ Rev.\ Lett. \textbf{81}, 5932--5935 (1998).

\bibitem{Bennett96}
C.~H. Bennett, G.~Brassard, S.~Popescu, B.~Schumacher, J.~A.
Smolin, and W.~K.
  Wootters, \enquote{Purification of Noisy Entanglement and Faithful
  Teleportation via Nois Channels,} Phys.\ Rev.\ Lett. \textbf{76}, 722--725
  (1996).

\bibitem{Pan03b}
J.-W. Pan, S.~Gasparoni, R.~Ursin, G.~Weihs, and A.~Zeilinger,
  \enquote{Experimental Entanglement Purification,} Nature \textbf{423},
  417--422 (2003).

\bibitem{Kaltenbaek2006}
R.~Kaltenbaek, B.~Blauensteiner, M.~Zukowski, M.~Aspelmeyer, and
A.~Zeilinger,
  \enquote{Experimental interference of independent photons,} Phys.\ Rev.\
  Lett. \textbf{96}, 0240,502 (2006).

\bibitem{Nordholt02}
J.~E. Nordholt, R.~Hughes, G.~L. Morgan, C.~G. Peterson, and
C.~C. Wipf,
  \enquote{Present and Future Free-Space Quantum Key Distribution,} in
  \emph{Free-Space Laser Communication Technologies {XIV}}, vol. 4635 of
  \emph{Proceedings of {SPIE}}, p. 116 (SPIE, 2002).

\bibitem{Rarity02}
J.~G. Rarity, P.~R. Tapster, P.~M. Gorman, and P.~Knight,
\enquote{Ground to
  satellite secure key exchange using quantum cryptography,} New Journal of
  Physics \textbf{4}, 82 (2002).

\bibitem{Aspelmeyer03b}
M.~Aspelmeyer, T.~Jennewein, M.~Pfennigbauer, W.~R. Leeb, and
A.~Zeilinger,
  \enquote{Long-Distance Quantum Communication With Entangled Photons Using
  Satellites,} IEEE\ J.\ Sel.\ Top.\ Quant.\ Elec. \textbf{9}, 1541--1551
  (2003).

\bibitem{Hughes02}
R.~J. Hughes, J.~E. Nordholt, D.~Derkacs, and C.~G. Peterson,
  \enquote{Practical free-space quantum key distribution over 10 km in daylight
  an at night,} N.\ J.\ Phys. \textbf{4}, 43 (2002).

\bibitem{Rarity_2001}
J.~Rarity, P.~Tapster, and P.~Gorman, \enquote{Secure free-space
key exchange
  to 1.9 km and beyond,} J.\ Mod.\ Opt. \textbf{48}, 1887--1901 (2001).

\bibitem{kurtsiefer02}
C.~Kurtsiefer, P.~Zarda, M.~Halder, H.~Weinfurter, P.~Gorma,
P.~Tapster, and
  J.~Rarity, \enquote{A Step Towards Global Key Distribution,} Nature
  \textbf{419}, 450 (2002).

\bibitem{Aspelmeyer03}
M.~Aspelmeyer, H.~R. B{\"o}hm, T.~Gyatso, T.~Jennewein,
R.~Kaltenbae,
  M.Lindenthal, G.~Molina-Terriza, A.~Poppe, K.~Resch, M.~Taraba, R.~Ursin,
  P.~Walther, and A.~Zeilinger, \enquote{Long-Distance Free-Space Distribution
  of Quantum Entanglement,} Science \textbf{301}, 621--623 (2003).

\bibitem{Resch05}
K.~Resch, M.~Lindenthal, B.~Blauensteiner, H.~B\"ohm, A.~Fedrizzi,
  C.~Kurtsiefer, A.~Poppe, T.~Schmitt-Manderbach, M.~Taraba, R.~Ursin,
  P.~Walther, H.~Weier, H.~Weinfurter, and A.~Zeilinger, \enquote{Distributing
  entanglement and single photons through an intra-city, free-space quantum
  channel,} Optics Express \textbf{13}, 202--209 (2005).

\bibitem{Peng05}
C.-Z. Peng, T.~Yang, X.-H. Bao, J.~Zhang, X.-M. Jin, F.-Y. Feng,
B.~Yang,
  J.~Yang, J.~Yin, Q.~Zhang, N.~Li, B.-L. Tian, and J.-W. Pan,
  \enquote{Experimental Free-Space Distribution of Entangled Photon Pairs Over
  13 km: Towards Satellite-Based Global Quantum Communication,} Phys.\ Rev.\
  Lett. \textbf{94}, 150,501 (2005).

\bibitem{Ursin_2006}
R.~Ursin, F.~Tiefenbacher, T.~Schmitt-Manderbach, H.~Weier,
T.~Scheidl,
  M.~Lindenthal, B.~Blauensteiner, T.~Jennewein, J.~Perdigues, P.~Trojek,
  B.~Oemer, M.~Fuerst, M.~Meyenburg, J.~Rarity, Z.~Sodnik, C.~Barbieri,
  H.~Weinfurter, and A.~Zeilinger, \enquote{Free-Space distribution of
  entanglement and single photons over 144 km,} quant-ph/0607182  (2006).

\bibitem{Pfennigbauer05}
M.~Pfennigbauer, W.~Leeb, G.~Neckamm, M.~Aspelmeyer, T.~Jennewein,
  F.~Tiefenbacher, A.~Zeilinger, G.~Baister, K.~Kudielka, T.~Dreischer, and
  H.~Weinfurter, \enquote{Accommodation of a quantum communication transceiver
  in an optical terminal (ACCOM),} Tech. Rep. ESTEC/Contract No.
  17766/03/NL/PM, European Space Agency (ESA) (2005).

\bibitem{Kaltenbaek04a}
R.~Kaltenbaek, M.~Aspelmeyer, T.~Jennewein, C.~Brukner,
M.~Pfennigbaue, W.~R.
  Leeb, and A.~Zeilinger, \enquote{Proof-of-Concept Experiments for Quantum
  Physics in Space,} in \emph{Quantum Communications and Quantum Imaging},
  R.~Meyers and Y.~Shih, eds., vol. 5161, pp. 252--268 (2003).

\bibitem{stuetz06}
M.~St\"utz, Master's thesis, University of Vienna (2006).

\end{thebibliography}


\newpage

\section*{List of Figure Captions}

Fig. 1. The dotted volume indicates the region where the
visibility $V$, detection efficiency $\eta$ and number of partners
$n$ allow for a multi-party quantum communication complexity
protocol which is more efficient than any classical one for the
same task. The volume corresponds to that given by
inequality~(\ref{inequality}).

\noindent Fig. 2. (Left) Setup demonstrating the photon
manipulation by a spatial light modulator (SLM)~\cite{stuetz06}.
The photon pairs produced by down conversion in a barium borate
(BBO crystal) are entangled in their orbital angular momentum,
represented by the Laguerre-Gaussian modefunctions. The mode
index corresponds to the orbital angular momentum of each photon.
The transformation between different modes is performed by
passing the photons through phase diffraction gratings containing
a phase singularity, which is generated by the SLM. (Right)
Demonstrating the transformation of the photon by the
computer-calculated hologram on the SLM. The coincidence between
the detectors $D0_A$ and $D1_B$ is shown. Due to the initial
correlation between the photons there are little coincidence
counts, unless the SLM performs a $-1$ transformation. This
clearly demonstrates that we are able to manipulate the orbital
angular momentum of the entangled photon by means of the
computer-generated hologram.

\noindent Fig. 3. Quantum communication links realized as uplinks
from ground to Space. Entangled photons are generated on ground,
and sent towards one or more Space-based receivers. If only one
receiver is available, link a) will allow single quantum
communication. If this were a GEO satellite, several
groundstations could see the very same receiver, for successive
quantum key exchanges (link d).  If a second receiver were
available, e.g. also in GEO (link b) or in LEO (link c), also the
study of fundamental aspects of quantum entanglement over large
distances may be accomplished.

\noindent Fig. 4. Key elements of the ground based source (Ground
Segement) and the satellite based receiver (Space Segment) for a
quantum communication uplink to satellites.


\newpage

\begin{table}[h!]
\begin{tabular}{|c|c|c|c|c|}
  \hline
  Outcome & Outcome & Basis & Error   \\
   Qubit 1 & Qubit 2 & Adaptation & Correction \\\hline\hline
  $s_{1}=0$ & $s_{2}=0$ & no: $B_{2}(\beta)$& no \\ \hline
  $s_{1}=0$ & $s_{2}=1$ & no: $B_{2}(\beta)$& $\sigma_{z}$  \\ \hline
  $s_{1}=1$ & $s_{2}=0$ & yes: $B_{2}(-\beta)$& $\sigma_{x}$  \\ \hline
  $s_{1}=1$ & $s_{2}=1$ & yes: $B_{2}(-\beta)$& $\sigma_{x}\sigma_{z}$  \\ \hline
\end{tabular}
\caption{Table 1}
\end{table}


  \begin{figure}[htbp]
  \centering
  \includegraphics[width=8.3cm]{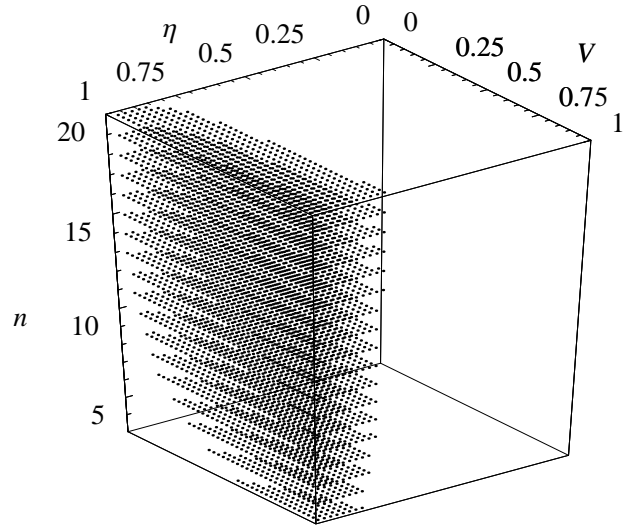}
  \caption{The dotted volume indicates the region where the
visibility $V$, detection efficiency $\eta$ and number of partners
$n$ allow for a multi-party quantum communication complexity
protocol which is more efficient than any classical one for the
same task. The volume corresponds to that given by
inequality~(\ref{inequality}).}
  \end{figure}
\newpage


  \begin{figure}[htbp]
  \centering
  \includegraphics[width=15.3cm]{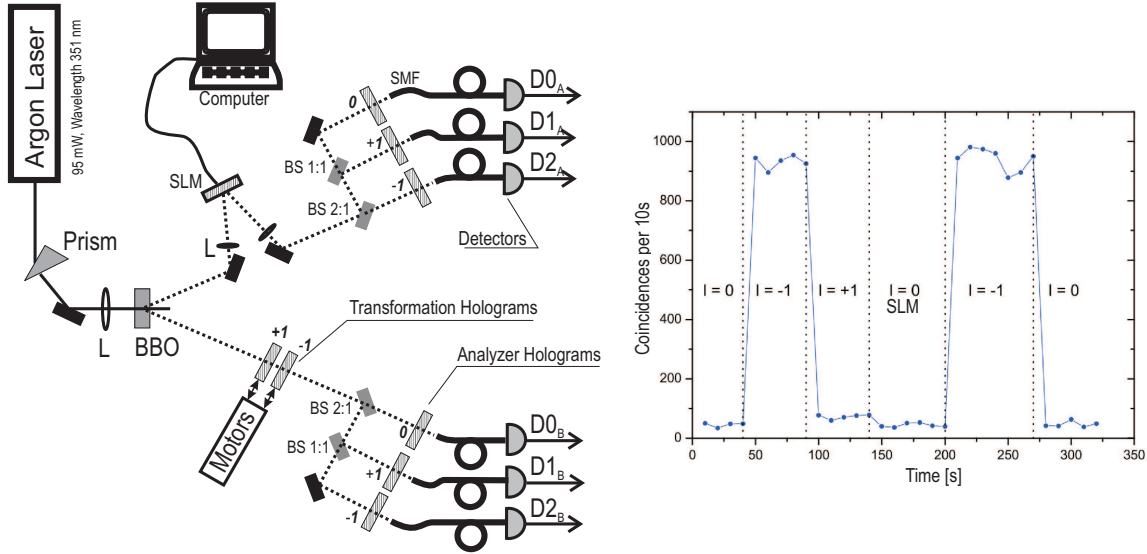}
  \caption{(Color online) (Left) Setup demonstrating the photon manipulation by a
spatial light modulator (SLM)~\cite{stuetz06}. The photon pairs
produced by down conversion in a barium borate (BBO crystal) are
entangled in their orbital angular momentum, represented by the
Laguerre-Gaussian modefunctions. The mode index corresponds to
the orbital angular momentum of each photon. The transformation
between different modes is performed by passing the photons
through phase diffraction gratings containing a phase
singularity, which is generated by the SLM. (Right) Demonstrating
the transformation of the photon by the computer-calculated
hologram on the SLM. The coincidence between the detectors $D0_A$
and $D1_B$ is shown. Due to the initial correlation between the
photons there are little coincidence counts, unless the SLM
performs a $-1$ transformation. This clearly demonstrates that we
are able to manipulate the orbital angular momentum of the
entangled photon by means of the computer-generated hologram.}
  \label{trits}
  \end{figure}

\newpage


  \begin{figure}[htbp]
  \centering
  \includegraphics[width=15.3cm]{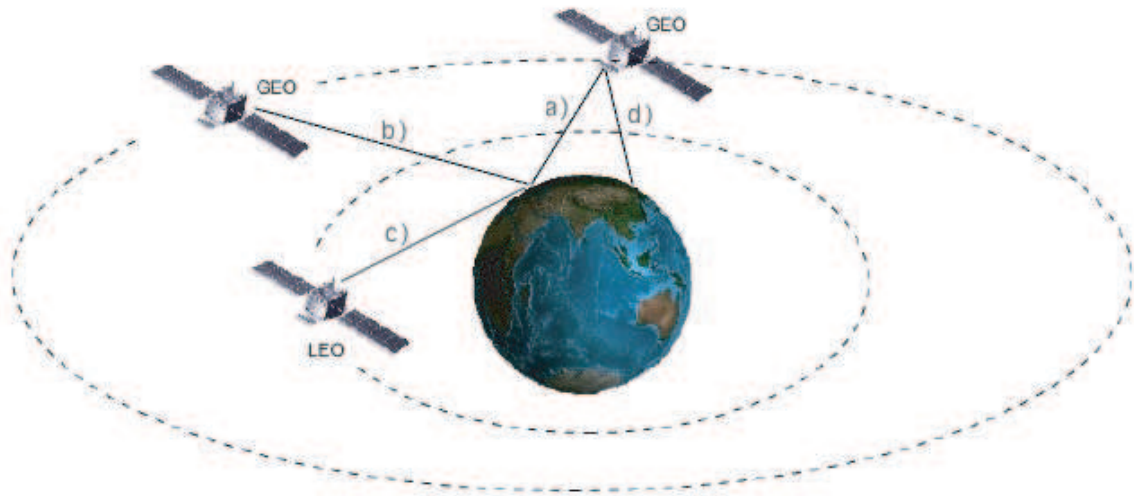}
  \caption{(Color online) Quantum communication links realized as uplinks from
ground to Space. Entangled photons are generated on ground, and
sent towards one or more Space-based receivers. If only one
receiver is available, link a) will allow single quantum
communication. If this were a GEO satellite, several
groundstations could see the very same receiver, for successive
quantum key exchanges (link d).  If a second receiver were
available, e.g. also in GEO (link b) or in LEO (link c), also the
study of fundamental aspects of quantum entanglement over large
distances may be accomplished.}
  \label{uplinks}
  \end{figure}

\newpage


  \begin{figure}[htbp]
  \centering
  \includegraphics[width=14.3cm]{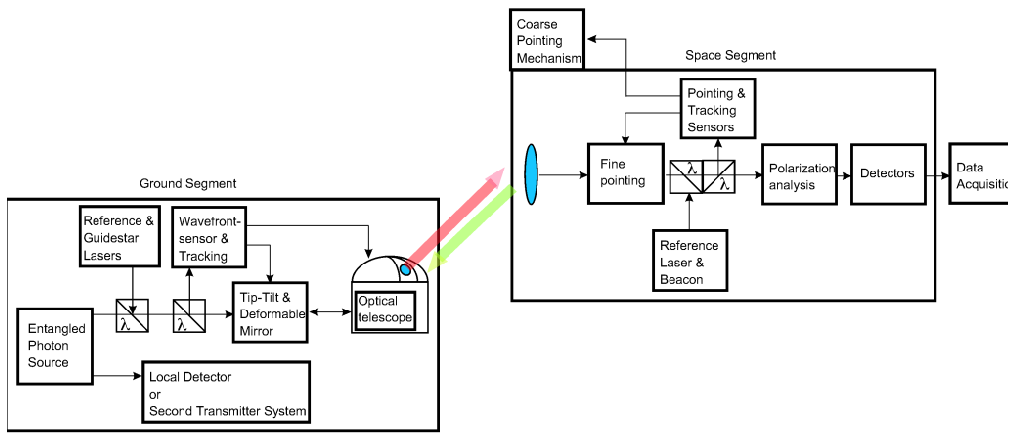}
  \caption{(Color online) Key elements of the ground based source (Ground
Segement) and the satellite based receiver (Space Segment) for a
quantum communication uplink to satellites. }
  \label{uplink_sys}
  \end{figure}

\end{document}